# Anomalous quantum metal in a 2D crystalline superconductor with electronic phase non-uniformity

Linjun Li[1,2,*], Chuan Chen[2,3], Kenji Watanabe[4], Takashi Taniguchi[4], Yi Zheng[5], Zhuan Xu[5],

Vitor M. Pereira[2,3,*], Kian Ping Loh[2,6,*], Antonio H. Castro Neto[2,3,*]

1 State Key Laboratory of Modern Optical Instrumentation, College of Optical Science and Engineering, Zhejiang University, Hangzhou, China 310027

2 Centre for Advanced 2D Materials and Graphene Research Centre, National University of Singapore, Singapore 117546

3 Department of Physics, National University of Singapore, 2 Science Drive 3, Singapore 117551

4 National Institute for Materials Science, Namiki 1-1, Tsukuba, Ibaraki 305-0044, Japan

5 Department of Physics, Zhejiang University, Hangzhou, China 310027

6 Department of Chemistry, National University of Singapore, 3 Science Drive 3, Singapore 117543

*Corresponding authors: lilinjun@zju.edu.cn, vpereira@nus.edu.sg, chmlohkp@nus.edu.sg, phycastr@nus.edu.sg.

**Abstract —** **The details of the superconducting to quantum metal transition (SQMT) at T=0 are an open problem that invokes much interest in the nature of this exotic and unexpected ground state[1-3]. However, the SQMT was not yet investigated in a crystalline 2D superconductor with coexisting and fluctuating quantum orders. Here, we report the observation of a SQMT in 2D ion-gel gated 1T-TiSe$_2$[4], driven by magnetic field. A field-induced crossover between Bose quantum metal and vortex quantum creeping with increasing field is observed. We discuss the interplay between superconducting and CDW fluctuations (discommensurations) and their relation to the anomalous quantum metal (AQM) phase. From our findings, gate-tunable 1T-TiSe$_2$ emerges as a privileged platform to scrutinize, in a controlled way, the details of the SQMT, the role of coexisting fluctuating orders and, ultimately, obtain a deeper understanding of the fate of superconductivity in strictly two-dimensional crystals near zero temperature.**

The superconductor-insulator transition (SIT) attracts considerable attention since it pitches the two extremes of electronic transport against each other: At a fundamental level, it deals with the contradicting facts that non-magnetic impurities are not expected to destroy superconductivity while, on the other hand, an electronic system with sufficient disorder is expected to be an insulator on account of Anderson localization[7]. In two-dimensional (2D) systems, the SIT has an increased complexity because those two extremes are peculiar: (i) any Fermi liquid metal in the presence of arbitrarily small disorder should be an Anderson insulator in the thermodynamic limit; (ii) the clean superconducting (SC) transition has an intrinsically topological nature[5] with correlations that decay only algebraically, as described by the Berezinskii-Kosterlitz-Thouless (BKT) theory, and the SC state is not expected to persist in non-zero magnetic fields (no Meissner phase). The SIT driven by magnetic field in the limit $T \rightarrow 0$ has been traditionally observed in disordered thin-film superconductors, and it has been considered the archetype of a quantum phase transition[2,3]. More recent experiments with cleaner systems have found that an anomalous quantum metal (AQM) generically intervenes the transition from the SC to the insulating phases, which has been attributed to a combination of disorder and quantum phase fluctuations[1-3, 6-13]. As a result, attention has recently shifted from the specifics of the SIT in "dirty" superconductors to the more encompassing problem of what kind of metallic state can ensue at $T \approx 0$ upon disrupting the SC order by magnetic field, how such a metal subsequently evolves into an insulator or a normal metal, and how the span of these regimes in the phase diagram of 2D superconductors is controlled by the amount of disorder.

So far, the AQM and the related nature of SC in strictly 2D systems has been scrutinized almost entirely in systems where spatial non-uniformity is of extrinsic origin. There are, however, a number of materials where SC coexists with other types of order – such as charge density waves (CDW) – characterized by spatially non-uniform textures that are intrinsic in nature. Yet, details of the SIT or the AQM state remain unexplored in such systems[14]. When such coexisting order is controllable with simple experimental parameters, it opens the door to an entirely new direction for exploring the interplay between superconductivity, disorder and quantum fluctuations because, ultimately, control over the coexisting order might allow control over the underlying non-uniformity and fluctuations. A specific example is the case of 1T-TiSe$_2$ nanosheets (TiSe$_2$, in short) in the atomically-thin 2D limit where the coexistence of intertwined SC and CDW domains is supported by X-ray diffraction experiments[15] and transport measurements in electrolyte-gated devices[4]. In the latter, flakes of TiSe$_2$ have been shown to have a $T$-vs-$n$ phase diagram globally in correspondence to that of bulk samples doped by Cu intercalation[16]. Namely, undoped 2D TiSe$_2$ undergoes a phase transition to a commensurate CDW phase (CCDW) at $T_{cdw} \sim 200$ K that persists down to zero temperature. SC emerges upon doping in a dome-shaped region beyond a critical density that coincides with the commensurate to near-commensurate transition of the underlying charge order where discommensurations proliferate[15, 17, 18]. This coincidence, whereby the SC order springs up precisely at the point where the long-range CDW phase coherence breaks up through discommensurations (phase slips), begs one to investigate the relation between the two orders. In addition to possibly driving or boosting SC pairing through fluctuation-assisted pairing, a network of these discommensurations provides a natural periodicity to potentially explain the observation of Little-Parks oscillations[4] which, in itself, offers evidence that the SC order in TiSe$_2$ develops at $T \lesssim T_C$ not in a uniform manner, but rather among an electronic matrix with a characteristic periodicity of the order of $\sim 100-500$ nm[4]. Hence, gate-tunable TiSe$_2$ crystals provide an ideal platform to investigate the nature of the SC state in a 2D material with distinct coexisting and fluctuating order parameters.

Here, we report the first observation and detailed study of the AQM in devices made with ion-gel-gated TiSe$_2$ nanosheets. Similarly to previous reports in a number of 2D systems[19], the AQM develops at low temperatures, $T < T_a$, where $T_a$ separates the thermal and quantum fluctuation regimes. By tuning the device to different densities around optimal doping we obtained the phase diagram presented in Fig. 1 in terms of temperature and magnetic field. At our lowest temperatures (0.25 K) and with the density tuned near the optimal doping (NOD), we find that the sheet resistance ($R_s$) scales with magnetic field as $R_s(H) \sim (H - H_{C0})^2$ for small field ($0.0\,\text{T} < \mu_0 H < 0.2\,\text{T}$), which then evolves to the exponential dependence characteristic of vortex quantum creeping closer to the upper critical field ($0.2\,\text{T} < \mu_0 H < H_{C2}$), in line with the expectation that higher magnetic field enhances quantum fluctuations at $T \approx 0$. In the following, in addition to describing these findings in detail, we conjecture on how the underlying near-commensurate CDW order might contribute to the AQM state in TiSe$_2$, which is an ingredient not present in other 2D superconductors studied so far.

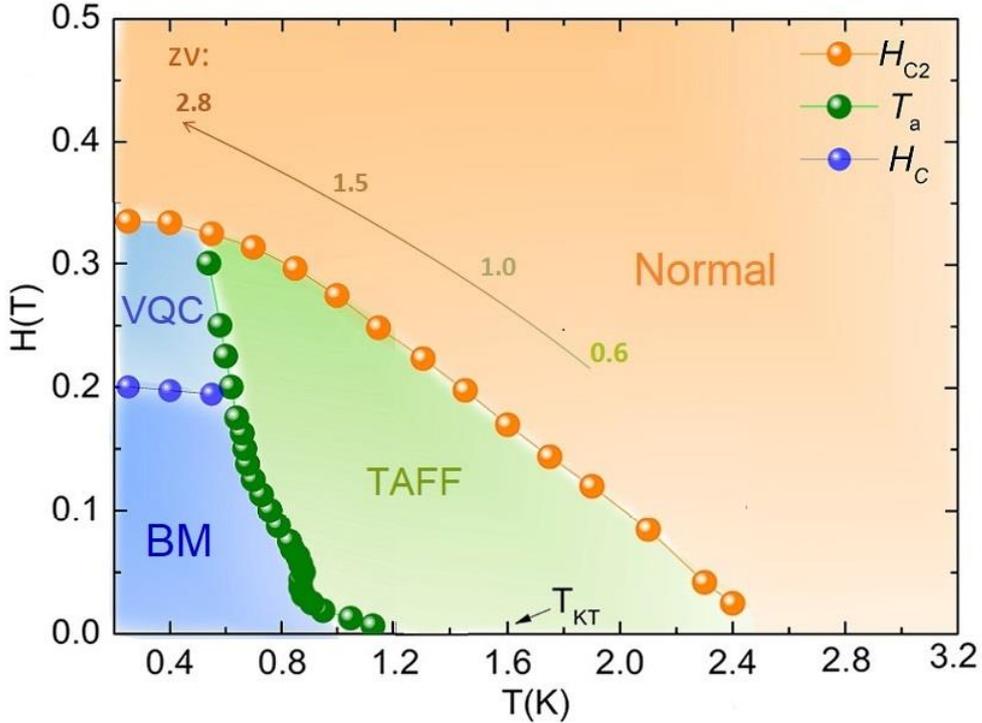

**Figure 1 – Phase diagram of two-dimensional superconducting 1T-TiSe$_2$.** Field-temperature phase diagram when the carrier density is tuned near to optimal doping ($n = 4.0 \times 10^{14}\,\text{cm}^{-2}$). Thermally assisted flux flow (TAFF) exists between the lines labeled $H_{C2}(T)$ and $T_a(H)$ while the anomalous quantum metal (AQM) is observed below the $T_a(H)$ line. With increasing field in the region $T \approx 0$, the system transitions from a dirty Bose metal (BM) regime

to vortex quantum creeping (VQC) around $H_C$, before reaching the normal state (Normal) for fields higher than $H_{C2}(T)$. The scaling exponent $zv$ obtained for different temperature ranges is also indicated. Its systematic increase from the clean to quantum percolation regimes with decreasing temperature is indicative of spatial inhomogeneity likely attributed to the underlying CDW order, and the dominant role of quantum fluctuations at the lowest temperatures (see text).

Our transport measurements were carried out in top-gated $TiSe_2$ electrical double-layer transistors (EDLT) using an ion-gel solution. As illustrated in Figs. 2(a,b), while the charge distribution in the gel is spatially uniform at zero gate (panel a), at finite gate voltages the electric field promotes charge separation and accumulation in surface charge layers that substantially contribute to increase the nominal carrier number in the target nanosheet of $TiSe_2$. This permits the large field-effect doping necessary to drive the system into the superconducting regime at low temperatures[20] and, as a result, allows one to map the phase diagram of $TiSe_2$ as a function of the key parameters shown in Fig. 1. As having the ion gel in direct contact with the $TiSe_2$ flake can lead to detrimental chemical reactions or ion intercalation, we separated them by encapsulating the whole device with an atomically thin spacer (1 to 2 layers) of crystalline hexagonal boron nitride. This is an important advantage in studying the possible existence of quantum phase transitions because it removes sample variability and the concomitant variation in the level of electronic disorder. A schematic of these devices is shown in Fig. 2 and additional details can be found in reference 4, which established this as a dependable and versatile strategy to map the phase diagram as a function of density using one single sample. To determine the device's resistivity, voltages were measured in a four-probe setup by lock-in techniques at low AC frequency (~ 17 Hz), combined with homemade frequency filtering to suppress high frequency noise (on the fridge at room temperature, with a cutoff frequency of 1 KHz), which is especially important to achieve thermalization at low temperatures[21].

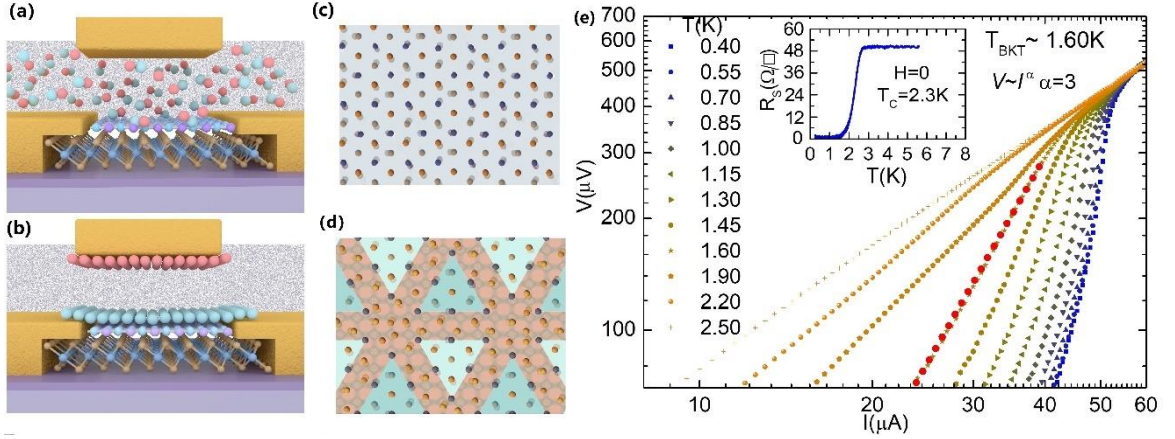

**Figure 2 – Two-dimensional superconductivity in a gated 1T-TiSe$_2$ nanosheet device.** (a) - (d) illustrate the process and consequences of ion gel gating. Without gate voltage (a), (c) the ion gel is not polarized and TiSe$_2$ undergoes only one phase transition as a function of temperature to its intrinsic, homogeneous, CDW phase. A finite voltage applied to the top gate (b), (d) polarizes the ion-gel which, in turn, dopes TiSe$_2$. In a finite interval about optimal doping and intermediate temperatures, the CDW phase breaks into commensurate domains separated by a network of discommensurations[44, 45]. Superconductivity emerges upon further lowering of temperature in this region. Panel (e) displays the measured *V-I* curves at NOD ($n = 4\times10^{14}$ cm$^{-2}$) and different temperatures, measured by lock-in techniques at 17 Hz. The transition temperature ($T_{BKT}$) is defined as that when $V \propto I^3$, and is highlighted in the plot by the red dashed line. The inset shows the resistive transition at zero field, where $T_C = 2.3$ K is defined as the temperature at which the resistance drops to 90% of the normal state value.

The measured device is 8 nm thick and has a width to length ratio $W/L = 6$, where *W* is the sample width and *L* is the separation between two voltage contacts. The zero temperature superconducting coherence length perpendicular to the plane is estimated to be $\xi_\perp = \xi_\parallel H_{c\perp}/H_{c\parallel} \approx 3.3$ nm, where $H_{c\parallel} \approx 3.3$T (derived from the in-plane magnetoresistance data shown in Supplementary Fig. 5), $H_{c\perp} \approx 0.35$ T (Fig. 1), and the in-plane coherence length is $\xi_\parallel = \sqrt{\phi_0/(2\pi H_{c\perp})} \approx 31$ nm. We note, however, that first-principles calculations[22] and other experiments[23] indicate that the carriers accumulated by electrolytic gating reside mostly in the layer of the sample closest to the ionic material, and the interlayer spacing in TiSe$_2$ is[24] c =

0.6 nm $\ll \xi_\perp$. Accordingly, we find the zero-field superconducting transition to be of the Berezinskii–Kosterlitz–Thouless (BKT) type which places our devices in the two-dimensional regime. Figure 2(e) shows the nonlinear *V-I* characteristics at different temperatures, when the carrier density determined from Hall measurements is tuned to $n \approx 4 \times 10^{14}$ cm$^{-2}$, near optimal doping (NOD) (the optimal doping level is around $5.9 \times 10^{14}$ cm$^{-2}$, as determined in ref. 4). The distinctive power-law behavior, $V \propto I^\alpha$, immediately below the critical current is characteristic of the BKT transition where the critical temperature $T_{BKT}$ is identified as that corresponding to α = 3. We obtain $T_{BKT} \approx 1.6$ K, indicated by the red dashed line in the figure. In addition, we define another temperature, $T_C$, for the onset of superconducting correlations as the temperature at which the resistance drops to 90% of the normal state value. For the NOD densities used in Fig. 2(c), we have $T_C \approx 2.3$ K, which is considerably above the BKT transition. On the one hand, a broad temperature separation between the onset of pairing at $T_C$ and the development of quasi long-range phase coherence at $T_{BKT}$ [ $(T_C - T_{BKT})/T_{BKT} \simeq 0.44$ ] in a clean system is expected as a result of the largely enhanced thermal fluctuations in this 2D crystal. On the other hand, such temperature separation could also be a signature of inhomogeneous superconductivity, in which case the transition width is proportional to the normal state resistance ($R_n$)[25]. In this context, we note that the device has $R_n \simeq 50$ Ω/□ at this doping, far from the resistance quantum, $R_Q \equiv h/4e^2 \simeq 6.4$ kΩ, and also much smaller than in typical MoGe and Ta thin films[26]. The $k_f l$ value is estimated as $k_f l = h/(2e^2)/R_n \approx 260$, much larger than the Ioffe-Regel limit ($k_f l \approx 1$), showing that the normal state is in the clean regime. Therefore, we attribute the relatively large separation between $T_C$ and $T_{BKT}$ here to strong thermal fluctuations in the phase of the superconducting order parameter, rather than extrinsic inhomogeneity. Importantly, being in the clean regime, enables the observation of the quantum metallic state in TiSe$_2$ at low temperatures, similarly to recent reports in crystalline bilayer NbSe$_2$[11] and ZrNCl[10].

It is noteworthy to mention that the characteristic nonlinear dependence $V \propto I^\alpha$, with α increasing monotonically from 1 at $\approx T_C$, passing through $\alpha = 3$ at $T_{BKT}$, and increasing further at lower temperatures, is maintained only at relatively large currents and voltages. As we show in the Supplementary Information, at voltages lower than those shown in Fig. 2(e), the V-I characteristics evolve to a linear dependence (i.e., $\alpha = 1$) at seemingly all temperatures below $T_{BKT}$. This contrasts with the nonlinear resistance of a conventional BKT transition in 2D superconductors[27-29] and will be revisited below.

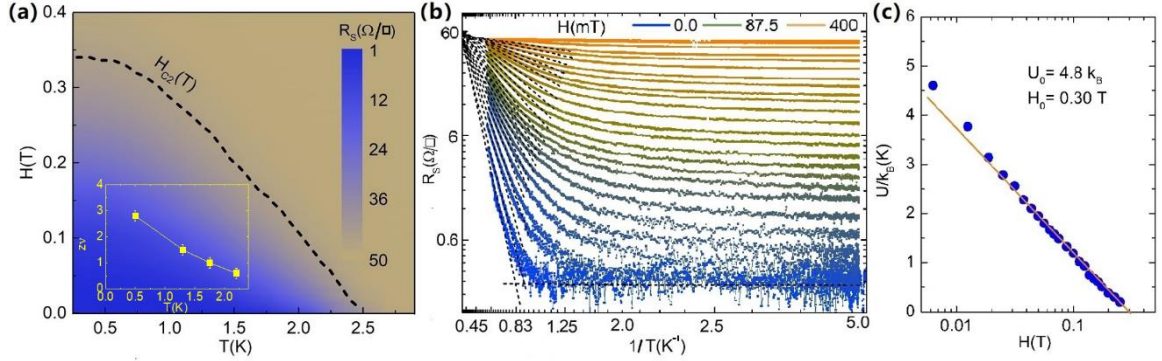

**Figure 3 – Magnetic field and temperature dependence of the resistance.** (a) Density plot of the sheet resistance ($R_S$) as a function of perpendicular magnetic field and temperature at NOD ($n = 4\times10^{14}$ cm$^{-2}$). The line labeled $H_{C2}(T)$ marks the onset of resistance drop with respect to the normal state. The inset shows the temperature-dependent scaling exponent $zv$ obtained as described in the text and further in the Supplementary Information. (b) The same data shown in (a), except that $R_S$ is now plotted against $1/T$ for different fields. The resistance drop is characterized by two distinct temperature regimes: at intermediate temperatures, below $T_C$, the resistance displays thermally activated behavior, as emphasized by the dashed lines in the figure. Below a field-dependent crossover temperature ($T_a$), $R_S$ plateaus at finite values establishing the presence of a metallic state at zero temperature. Panel (c) displays the semi-logarithmic plot of $U/k_B$ vs $H$, where $U$ is the thermal activation energy derived from the slope of the dotted lines in (b). The data can be fit with a dependence $U(H) = U_0 \ln(H_0/H)$ (solid line).

Figure 3 summarizes the magnetic field dependence of the sheet resistance at NOD. Panel (a) shows the resistance as a function of both temperature and the perpendicular external field, $H$. If we define a temperature-dependent upper critical field, $H_{C2}(T)$, as the threshold at which the resistance crosses 90% of the normal state value, we see that it displays the typical mean-field-like diagram (dashed line in Fig.3a). Constant-field traces are plotted in panel (b) as a function of reciprocal temperature and show the hallmarks of the AQM state in the extrapolated $T \approx 0$ limit (the lowest achievable temperature in our experiments is 0.25 K): the saturation of the sheet resistance, $R_S(T\approx 0)$, at finite values much smaller than $R_n$, combined with the systematic increase of $R_S(T\approx 0)$ with increasing field up to $H_{C2}(0)$. There is a giant magnetoresistance in this regime, with the resistance at saturation spanning close to three orders of magnitude below $R_n$.

The dissipative regime that takes place between the onset of resistance saturation and the normal state shows behavior characteristic of thermally activated flux (vortex) flow (TAFF): $R_S(T, H) = R_0 \exp[-U(H)/k_B T]$, which is highlighted by the dashed lines in Fig. 3b. The activation barrier is seen in Fig. 3c to vary with magnetic field as $U(H) = U_0 \ln(H_0/H)$, which is typical of collective flux creeping[1, 30], and our fit yields $U_0 = 4.8$ K and $H_0 = 0.30$ T. In order to mark the crossover between the TAFF and the AQM regimes for a given field, we define $T_a$ as the temperature at which the dashed line (TAFF) intersects the solid horizontal one (saturation); its dependence on magnetic field is shown in Fig. 1, where we see that $T_a$ decreases with increasing field.

The nature of the AQM state and the physics underlying the SQMT has been a vigorously explored topic ever since experiments demonstrated that disordered thin superconducting films can be driven across a superconductor-insulator transition (SIT), either by varying disorder or magnetic field at $T\rightarrow 0$[31, 32]. In the case of the field-tuned transition, it was initially interpreted as a manifestation of a quantum critical point separating a vortex glass phase (superconducting due to vortex pinning) from a Fermi glass (insulating due to the low dimensionality and strong disorder), since the resistivity data obeyed the theoretically predicted universal scaling for the quantum critical point in that scenario[33, 34]. According to this picture, at $T = 0$ there would be either a true zero-resistance state ($R_s = 0$) at small magnetic fields or an insulating state ($R_s = \infty$) at higher fields, in a perfect contrast between superconductivity and Anderson localization. A finite resistance at $T = 0$ would only obtain at the critical field[35] and should have a universal value $\sim h/4e^2$. This scenario was brought into question by the realization that there was no strict universality[26]. More precisely, it was subsequently established that the SIT is seemingly unique to strongly disordered systems, and that a finite, field-dependent $R_s$ ubiquitously obtains as $T \rightarrow 0$ in relatively clean samples characterized by $R_n \ll R_Q$[19].

That an AQM state like the one we characterize in Fig. 3 should exist at $T=0$ has challenged the microscopic understanding of the possible ground states in this problem[9]. On the one hand, the fact that $R_s \ll R_Q$ combined with (i) a large magnetoresistance, (ii) nonlinear I-V curves, (iii) the vanishing of Hall resistivity[36], (iv) the absence of Drude dynamics[37], and (v) values of $T_a$ typically much smaller than $T_C$, collectively suggest that the key physics is enacted by Cooper pairs with no phase coherence (superconducting fluctuations). As such, the dominant physics at play should be that of interacting charged bosons in the presence of disorder (the so-called dirty boson picture[33]). The difficulty, however, has been that a metallic phase in such a system is a non-trivial outcome because the conventional understanding of dirty bosons at $T=0$ only encompasses a superconducting (superfluid) to insulating phase transition. As a result, the quest for the ultimate nature of this Bose metal has become a fundamentally interesting problem in condensed matter physics. Ideas based on vortex liquid states[4], quantum tunneling of vortices in a vortex glass[12], inhomogeneous weak superconductivity[38], or a two-fluid formulation incorporating fermionic quasiparticles in a model of dirty bosons[39] have been advanced. One particular class of models explores the role played by coupling of the superconducting phase fluctuations to quenched disorder and dissipation mechanisms[2,6,7,40]. Das and Doniach[6,7] on the one hand, and Dalidovich, Wu and Phillips[8,41] on the other, have explored the possible emergence of a Bose metal in the so-called phase glass at finite fields. In particular, both groups predicted the resistance at the zero-temperature SQMT to scale with field as

$$R(H) \propto (H - H_{C0})^y, \qquad (1)$$

where $H_{C0}$ is the critical field and $y$ is derived from the power-law divergence of the superconducting coherence length[7,41]; Wu and Phillips specifically predict $y=2$. This scaling has been observed recently[11] in the 2D superconducting bilayer $NbSe_2$ and, in the following, we describe how this is also observed at the SQMT in our case of $TiSe_2$ at small fields.

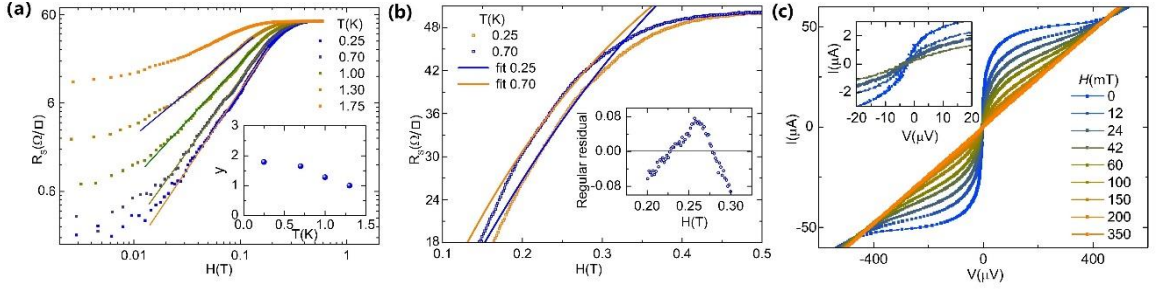

**Figure 4 – Magnetic field scaling of the resistance in the anomalous metallic state.** (a) In a field range $H_{C0} < H < H_C$, where $H_C < H_{C2}$ is a crossover field that depends weakly on temperature (see Fig. 1a), the sheet resistance at fixed $T$ can be described by the dependence $R_S \propto (H - H_{C0})^y$ characteristic of the phase glass model in the low field limit. The inset reports the exponent $y$ obtained by fitting the curves at the different temperatures shown in the main graph. Panel (b) shows that, at higher magnetic fields, $H_C < H < H_{C2}$, the measured resistance can be satisfactorily fit by the expression in equation (2) predicted for dissipation due to quantum creeping of vortices. (c) The magnetic field dependence of the *I-V* curves at $T = 0.25$ K. The inset magnifies the behavior near the origin at small fields where no hysteresis is observed.

Fig. 4a reports the field dependence of the longitudinal resistance at temperatures below $T_a$ (i.e., in the saturation region of Fig. 3b, outside the TAFF regime). There is an interval approximately between 0.01 T and 0.2 T where it behaves as $R(H) \propto (H - H_{C0})^y$ (below 0.01 T, the data falls to noise level), as highlighted by the straight lines superimposed on each experimental curve. The exponent $y$ that best fits the data in this interval of magnetic field is indicated in the inset; it approaches $y \approx 2$ at our lowest temperature, which tallies with the calculations of Wu and Phillips[41]. The fitting also gives us $H_{C0} \approx 0$ T at the lowest temperature, which is in accordance with the fact that the $R(H)$ traces in Fig. 4a have an upward-positive curvature as $H \to 0$, indicative of the finite resistance at zero field all the way down to, and including, $T = 0.25$ K (but recall also our earlier remarks related to the noise level in our experiments).

This scaling is no longer capable of accurately describing the measured traces above $H\approx 0.2\,\text{T}$ and up to $H_{C2}$, as is clear already in Fig. 4a from the negative curvature of the data in that field range. This discrepancy is shown in more detail in the supplementary Fig. S2a. Instead, we found that the resistance is best described at these fields by the dependence proposed for dissipation in superconductors due to quantum tunneling of vortices (quantum creeping)[13,15] at $T<T_a$:

$$R(H) \propto R_Q \frac{\kappa}{1-\kappa}, \quad \kappa \propto \exp\left[C\pi(\frac{h}{R_n e^2})\frac{H-H_C}{H}\right], \qquad (2)$$

where $C$ is a dimensionless constant. Fig. 4b shows the best fit to this field dependence. The AQM in our TiSe$_2$ sample is thus exhibiting a crossover at $H \sim H_C$ between the regimes described by equations (1) and (2). The obtained crossover field, $H_C$, is essentially independent of temperature (Fig. 1a), as expected from a purely quantum-mechanical mechanism. Such specific crossover has been predicted by Das and Doniach, albeit in the context of strongly disordered superconductors where the AQM gives way to an insulating state at high field[7]. Nevertheless, despite our TiSe$_2$ sample being in the clean region, one expects the same physics underlying the field-driven crossover between the two regimes to be at play here. Namely, whereas in the phase glass picture the SQMT and the scaling (1) is primarily a result of gauge field fluctuations[7,41], these are overcome by zero-point quantum fluctuations at higher fields. In this case, the resistance becomes "activated" with magnetic field which is the parameter controlling the strength of quantum fluctuations[7,12]. While the field-dependence (1) has been reported recently in NbSe$_2$[11] and the behavior (2) has been separately seen in ZrNCl[10], the observation of the crossover in a single sample is, to the best of our knowledge, so far unique to gate-doped TiSe$_2$.

In Fig. 4c, we plot the four probe $I$-$V$ curves at different magnetic field. As the field is reduced from $H_{C2}$, the progressively steeper slope at the origin indicates the transition from the normal state to the AQM. It is worth remarking that, unlike the case reported in monolayer NbSe$_2$[11], we see no hysteresis in the $I$-$V$ curves down to the smallest fields, which is consistent with the above analysis that estimates $H_{C0} \approx 0\,\text{T}$.

Our observations related to this TiSe$_2$ device with density tuned at NOD are globally summarized in the *H-T* phase diagram of Fig. 1. The normal state is separated from the phases exhibiting superconducting correlations at low *T* and low *H* by the mean-field-type $H_{C2}(T)$ line. At zero field, superconducting correlations begin developing at $T_C \approx 2.3$ K, but vortex-antivortex excitations remain unbound down to $T_{BKT} \approx 1.6$ K, at which temperature the BKT transition takes place and a true superconducting state ($R_s = 0$) is expected in a prefect 2D superconductor. At finite but small fields, the system transitions from the normal state to the TAFF regime at $T_C(H)$ with decreasing temperature, followed by a crossover to the AQM regime below $T_a$. In the portion of the phase diagram below the line $T_a(H)$ it remains metallic as $T \rightarrow 0$, although the extrapolated sheet resistance can be more than two orders of magnitude smaller than $R_n$ (Fig. 3b and Supplementary Fig. S1); this giant positive magnetoresistance is in correspondence with the prevalent behavior of the AQM in a variety of other 2D superconductors having $R_n \ll R_Q$[19]. With increasing field, the resistance of the AQM displays the crossover discussed above which, although first predicted in reference 10, has remained unreported and can be an indication that, in TiSe$_2$, there is a more substantial enhancement of quantum fluctuations by the external field.

There are two important issues that we should now briefly address: dissipation and spatial non-uniformity. In relation to the first, one must ponder whether the AQM might be simply an effect arising from the presence of the ionic gate placed at a sub-nanometric separation from the current channel, thus acting as a potential source of dissipation. However, controlled experiments in MoGe$_x$ indicate that the close proximity of a metallic gate tends to act *against* the metallic state[42]. This aspect has been more clearly reinforced in a recent experiment on NbSe$_2$ flakes that, by deliberately varying the level of external dissipation, shows the latter tends to stabilize the zero-resistance, superconducting state[43]. This, combined with the similarity in the behavior seen here in TiSe$_2$ to that reported in other ionic-gated crystalline superconductors[10], supports an intrinsic origin of the AQM state. At the level of microscopic models, metallic states have certainly been obtained when dirty-bosons are coupled *ad-hoc* to gapless degrees of freedom. Nonetheless, Phillips and collaborators have shown that the Bose-metallic state does not necessarily require dissipation but only a lack of phase coherence[7, 8] (plus interactions) and, moreover, have shown that dissipation can in fact be self-generated by coupling to the gapless excitations of the phase glass[41] (it is in the context of this model that their specific prediction $y=2$ for the scaling in equation (2) arises). For the case under consideration, however, it is important to recall that the superconducting dome of TiSe$_2$ arises from (and coexists with) a well-established CDW order. Crucially, the fact that superconductivity emerges only when doping causes a commensurate to near-commensurate CDW transition[15, 17, 18] is unlikely to be a coincidence[44]: The presence of CDW phase fluctuations can simultaneously contribute to enhance the pairing (by fluctuation-induced pairing, thereby explaining the coincidence in the loss of CDW commensuration with the superconducting dome) as well as provide a natural, intrinsic dissipation channel for the preformed Cooper pairs necessary to stabilize the Bose-metal.

This brings us to the issue of spatial non-uniformity. Irrespective of whether CDW fluctuations conspire or not to promote superconducting pairing and dissipation, their presence unavoidably implies an intrinsic non-uniformity of the electronic system. In fact, it is known that the mean-field-level solution of the commensurate to near-commensurate CDW transition in this class of dichalcogenides consists, in the vicinity of the transition, of a 2D superlattice structure of finite-sized CCDW domains separated by domain boundaries in the form of phase slips of the density order parameter[44, 45]. Since STM measurements have shown that this state of affairs remains below the superconducting $T_C$[4], we conjecture that the spatial non-uniformity of the underlying electronic system in the normal state translates into the non-uniform development of a superconducting order parameter[4, 44]. Ironically, despite its clean crystalline nature, this would render the microscopic situation in TiSe$_2$ somewhat similar to that of disordered superconducting films, although for very different reasons and with important differences: (i) the inhomogeneity is intrinsic, rather than extrinsic; (ii) it is not caused by static disorder, but rather a combination of a large-scale mean-field (static) super-periodicity of mostly commensurate CDW domains with dynamical CDW fluctuations. In this situation, a natural starting point is to consider a network of superconducting domains Josephson-coupled to each other, which indeed has been the common starting point in most attempts to model the Bose metal microscopically. Since the CDW does not gap the electronic spectrum and TiSe$_2$ is thus a relatively good metal in the non-superconducting NCCDW state, it is reasonable to assume that, in the above "granular" picture, the superconducting domains are embedded in a metallic matrix alive with CDW fluctuations (indirect evidence for this metallic background is provided by the presence of a zero-bias conductance peak in these devices[4]). The conditions for the development of an AQM phase in this scenario have been studied in detail by Spivak and collaborators[46] (although without any coexisting CDW order).

Despite the absence of a definite quantum critical point in systems hosting the AQM phase[2], in an attempt to characterize the spatial inhomogeneity, we performed a scaling analysis of the $R_s(T,H)$ traces in different temperature regions, whose details can be found in part 3 of the supporting information. The values obtained for the dynamic critical exponent $zv$ (with fitting error of about ±0.2) are reported in the phase diagram of Fig. 1a, as well as plotted in the inset of Fig. 3(a). At high temperatures, near or above $T_{BKT}$, we obtain $zv \sim 0.6-1$ which is characteristic of the clean limit in the universality class of the 2+1 D $XY$ model[37] ($zv = 2/3$). As the temperature is lowered, thermal fluctuations are suppressed and the SC order is likely to be gradually stabilized within the incommensurate CDW matrix. The value of $zv$ is higher at lower temperatures, consistent with quantum fluctuations progressively taking over thermal ones: near $T_a$, $zv$ is close to the value characteristic of a classical percolative transition ($zv = 4/3$) while at the lowest temperatures (below $T_a$) it evolves to that of quantum percolation ($zv = 7/3$). This monotonic increase of $zv$ with decreasing temperature could be an indication of proximity to a quantum Griffiths state[47]. Recent experiments on Ga thin films[48], LaAlO$_3$/SrTiO$_3$ interfaces[49], as well as ZrNCl and MoS$_2$ thin films[50] have observed similar behavior of the dynamical exponent, including a strong divergence at a characteristic magnetic field and extremely low temperatures (typically a few dozen mK), which is the hallmark of the quantum Griffiths critical point[47]. In our case, we do not detect such singular behavior in $zv$, nor an upturn in the critical $H_{c2}(T)$[50]. However, since extremely low temperatures seem to be required to probe the vicinity of this critical point[48], but our setup is limited to a minimum of 0.25 K, it is unclear at this stage whether the trend in $zv$ is here also a result of underlying quantum Griffiths physics, or simply a manifestation of cross-over between different regimes associated with with finite $zv$ due to an underlying microscopic inhomogeneity.

We now discuss the phase diagram at $H=0$ when $T<T_{BKT}$. We note that the zero-field trace of $R_s(T)$ in Fig. 3b still displays a knee at $\sim T_a$ which foreshadows the saturation of resistance as $T \to 0$. This raises the question of whether two-dimensional TiSe$_2$ crystals harbor a truly superconducting state ($R_s = 0$) in zero field. On the one hand, having $R_s \neq 0$ down to the smallest magnetic fields probed in our experiment is in line with the expectation that $H_{C1} = 0$ in a strictly 2D and strictly clean system (i.e., one where there is no disorder to pin vortex motion under arbitrarily low field below $T_{BKT}$)[30]. Still, if $H=0$, under normal circumstances one would expect uniform superconductivity below the vortex unbinding temperature, $T_{BKT}$. But the presence of the underlying CDW fluctuations is likely to change this, possibly to such an extent that these quantum fluctuations might break long-range phase coherence and quench the true superconducting state, even without applying a magnetic field. If that were the case, the addition of an external field should further boost the effects of quantum fluctuations and one might expect a strong field sensitivity. In this regard, it is noteworthy that, when compared to the cases of ZrNCl and NbSe$_2$, the value of $H_{C2}$ in our 2D samples of TiSe$_2$ is considerably smaller than the corresponding critical field in the bulk[16]. As the lower dimensionality generically magnifies quantum fluctuations, this qualitatively agrees with the view that the superconducting order in our 2D devices is already strongly affected by CDW fluctuations and, hence, less field is required to break the superconducting correlations. While our current measurements are not conclusive, the results reported here strongly encourage further scrutiny of the interplay between these two coexisting and fluctuating orders.

Finally, we consider extrinsic factors that might account for the AQM state observed in our experiments. Generically, these could include measurement artifacts, failure in cooling or thermally equilibrating the electronic system due to Joule heating or electrical radiation noise[21], coupling to dissipative degrees of freedom[2, 42], or finite size effects[51]. The behavior reported here has been observed consistently in all three measured devices prepared under the same conditions, and at different densities within the superconducting dome (see Supplementary Fig. S6). In addition to the low normal-state resistance, atomic force microscopy scans taken on the device surface reveal a characteristic roughness of only ~ 1 nm, which is far less than the typical surface roughness of amorphous MoGe films, and further reinforces the uniform nature of our sample and device. Arguments similar to those advanced in the context of the AQM in MoGe[1] would in principle exclude the possibility of incomplete cooling in our experiment. In addition, as highlighted earlier, our transport setup deliberately included low-pass frequency filters as a precaution against exposure to unwanted radiation at low temperatures[21]. Yet, we pointed out earlier that, as shown in Supplementary Fig. S4, the zero-field voltage crosses over from a nonlinear ($V \propto I^{\alpha}$) back to a linear current dependence at low current and $T < T_{\text{BKT}}$. Calculations have shown that such crossover can arise as a manifestation of finite-size effects related to the device dimensions or, more generically, when the system harbors any length scale that cuts-off the logarithmic vortex interaction that is necessary to ensure the complete BKT transition[51]. In the case of TiSe$_2$, such a length scale arises naturally, and intrinsically, from the network of CDW discommensurations that is expected to be present at the densities where the SC dome is stabilized[18, 44], and for which there is evidence from STM measurements in the SC phase[52]. Whether or not this ultimately explains the absence of a zero-resistance state at $H = 0$ in our experiment, and whether the dissipative state is caused by the length scale associated with the underlying CDW texture or simply a matter of device geometry and dimensions will require further scrutiny across more samples and devices. In this context, the fact that doping in EDLTs is likely effective only at the topmost layer of the TiSe$_2$ film, requires the magnetotransport characterization of a strict monolayer crystal to avoid ambiguities related to the possible coexistence of superconducting and metallic monolayers in few-layer films such as those used here.

In conclusion, we have made detailed magnetotransport measurements on an ion-gel gated 2D device to explore the nature of superconductivity in TiSe$_2$ sheets as a function of magnetic field and deviations of density from optimal doping. Having a BN spacer and a single-crystal sample ensures a clean device. Our key observations are summarized in the phase diagram of

Fig. 1. Most notably, between our lowest temperatures and $T_a \sim 0.7$–$1.0$ K, transport is dominated by an anomalous quantum-metallic phase at all finite fields. The giant positive magnetoresistance in this phase displays a crossover between the two regimes proposed within the phase glass picture[6-8, 41], which are designated "Bose metal" and "vortex quantum creeping" in the phase diagrams of Fig. 1. Since the onset of SC behavior in TiSe$_2$ coincides with the disruption of commensurate CDW order through discommensurations, we advance that the development of the SC order parameter is inescapably intertwined with that of the charge density and its fluctuations. This has a direct implication in terms of providing both an intrinsic spatial non-uniformity for the development of the AQM, as well as a natural dissipation channel via phase fluctuations of the CDW. The presence of an additional quantum-fluctuating order parameter might explain what, from our data, seems to be a persistence of the AQM in the absence of magnetic field which suggests the absence of a truly zero-resistance state in TiSe$_2$. These findings and the gate-tunability of TiSe$_2$ open the door to exploring, in a controlled way, the fate of superconductivity in 2D in the presence of competing or coexisting orders.

**Acknowledgments** – We acknowledge fruitful discussions with Lei Su on the interplay between superconductivity and CDW order in TiSe$_2$, and Eoin O'Farrell and Wei Fu in preparing the manuscript. Li Linjun acknowledges the funding support from the Chinese government "one thousand young talents" program and a grant from the National Natural Science Foundation of China (No.11774308). Xu Zhuan acknowledges The National Key Projects for Research & Development of China (Grant. No. 2016FYA0300402).

**Author contributions:** L. J. L. carried out the experiment under the supervision of K. P. Loh. L. J. L., C. C, A. H. C. N, and V. M. P. analyzed the data. L. J. L and V. M. P wrote the manuscript. K. W, T. T provided the hBN single crystals. All authors reviewed and commented on the manuscript.

**Conflict of interests:** The authors declare no competing financial interests. Requests for materials should be addressed to L. J. L. (lilinjun@zju.edu.cn).

# Supporting Information

The Supporting Information is available free of charge on the ACS Publications website at DOI:

(Density dependence of the superconducting transition and periodic magnetoresistance oscillation, fitting quality of Bose metal and vortex quantum creeping models, the scaling of magnetic field dependence of the resistance, V-I characteristics at different temperatures, estimation of the in-plane upper critical field, reproducibility across different devices and densities.)

# Supporting information

# Anomalous quantum metal in a 2D crystalline superconductor with electronic phase non-uniformity

Linjun Li[1,2,*], Chuan Chen[2,3], Kenji Watanabe[4], Takashi Taniguchi[4], Yi Zheng[5], Zhuan Xu[5],

Vitor M. Pereira[2,3,*], Kian Ping Loh[2,6,*], Antonio H. Castro Neto[2,3,*]

1 State Key Laboratory of Modern Optical Instrumentation, College of Optical Science and Engineering, Zhejiang University, Hangzhou, China 310027

2 Centre for Advanced 2D Materials and Graphene Research Centre, National University of Singapore, Singapore 117546

3 Department of Physics, National University of Singapore, 2 Science Drive 3, Singapore 117551

4 National Institute for Materials Science, Namiki 1-1, Tsukuba, Ibaraki 305-0044, Japan

5 Department of Physics, Zhejiang University, Hangzhou, China 310027

6 Department of Chemistry, National University of Singapore, 3 Science Drive 3, Singapore 117543

*Corresponding authors: lilinjun@zju.edu.cn, vpereira@nus.edu.sg , chmlohkp@nus.edu.sg, phycastr@nus.edu.sg.

# 1. Density dependence of the superconducting transition and periodic magnetoresistance oscillation

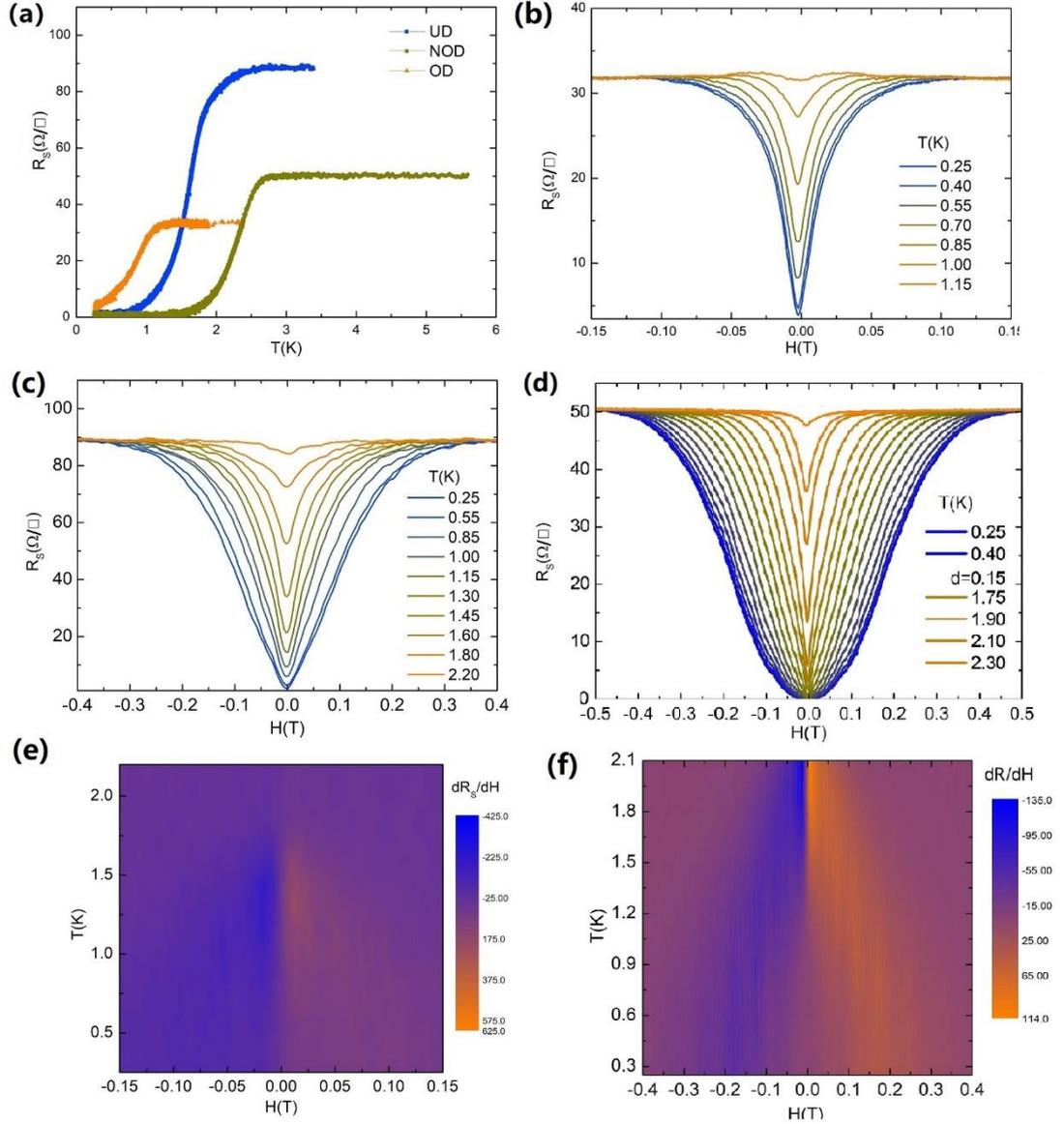

**Figure S1** – (a) The zero-field superconducting transition of three different doping levels: underdoped (UD), near optimum doped (NOD) and overdoped (OD). (b) Field-dependent magnetoresistance at different temperatures for OD sample. (c), (d) Field dependent *MR* at different temperatures for the UD and NOD samples respectively. (e), (f) Density plots of d$R_s$/d$H$ as a function of field and temperature, obtained from the traces in (c) and (d),

respectively. MR oscillations periodic in magnetic field can be distinguished in the faint vertical stripes of each color map.

Three different doping levels superconducting transition are present here: underdoped (UD: 2.1 $\times 10^{14}$ cm$^{-2}$), near optimum doped(NOD: $4\times 10^{14}$ cm$^{-2}$) and highly overdoped (OD: $12\times 10^{14}$ cm$^{-2}$). The zero-field superconducting transition and field dependent magneto-resistances at different temperatures for all three dope levels are displayed in Fig.S1. The upper critical field $H_{C2}$ values plotted in the phase diagram Fig.4a are derived by the definition of the resistance point of 90% normal state resistance, correspondingly the 90% of the normal state voltage. In Fig.S1(f), we show the 2D contour plot of the dR/dH data versus temperature and magnetic field for NOD. One can see, similar to our previous observed, the dR/dH values are periodically oscillating with magnetic field. The magnetic field periodicity is about 0.05T. The zero-resistance superconducting state(ZRSC) is determined by the flattened voltage level at the resolution limit of the detecting equipment, lock-in amplifier (Stanford Research SR830), normally about 10 nV. The excitation current of 100nA and the a.c. frequency of 13.373Hz are used in the measurement of magnetoresistance. The I-V curves are measured by using a combined DC setup of Keithley 2400 and Keithley 2000.

## 2. The fitting quality of Bose metal model and vortex quantum creeping model

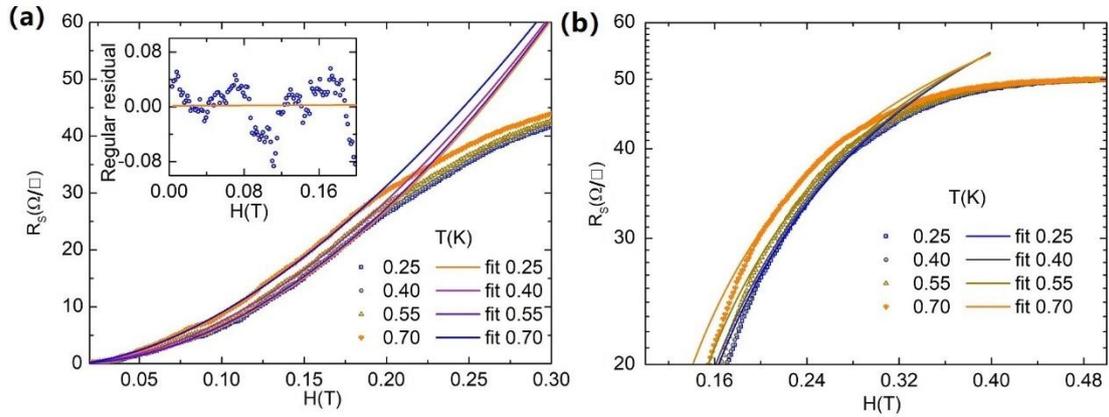

**Figure S2** – (a) Fitting of the field dependence of the sheet resistance to the Bose metal scaling in the low field region. The inset displays the regular residual of the fit which reflects the fitting quality. (b) Fitting of the same data to the vortex quantum creeping model in the high field region.

In Fig. S2, we show the linear plot of $R_{xx}$ vs $H$ for both fittings of Bose metal model and vortex quantum creeping model. One can see the fitting are both fine in the respective fitting range, with the regular residual distribution around zero in balance, which is shown in the inset plot of both figures.

## 3. The scaling of magnetic field dependence of resistance in different temperatures

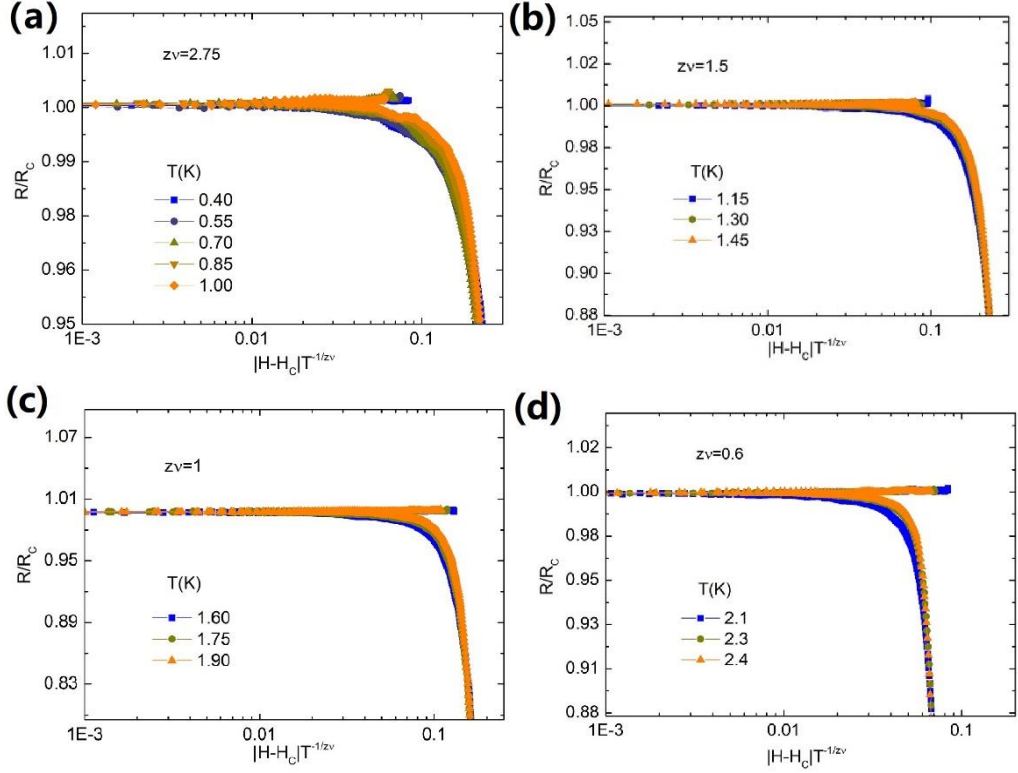

**Figure S3** – Fitting of the magnetoresistance to the critical scaling in small intervals centered at different temperatures. From (a) to (d), the reference temperature increases and the vz value decreases.

By using the scaling formula $R_S/R_C = F((H - H_C) \times T^{-1/vz})$, where $R_C$, $H_C$ are two fitting parameters, $F$ is an arbitrary function with $F(0) = 1$, the data are expected to collapse into two sets of lines, with a certain $vz$ value. As can be seen in Fig.S3, from a to d, that is from low temperatures to high temperatures, the fitted $vz$ values are decreasing from 2.75 to 0.6, which indicates that the system experiences three stages of clean limit ($vz = 2/3$), classical percolation ($vz = 4/3$), quantum percolation ($vz = 7/3$).  The increase in vz at smaller temperatures suggests that the system evolves from a clean-like to dirty-like regime in the language of disordered superconductors.

## 4. V-I characteristics as a function of temperature

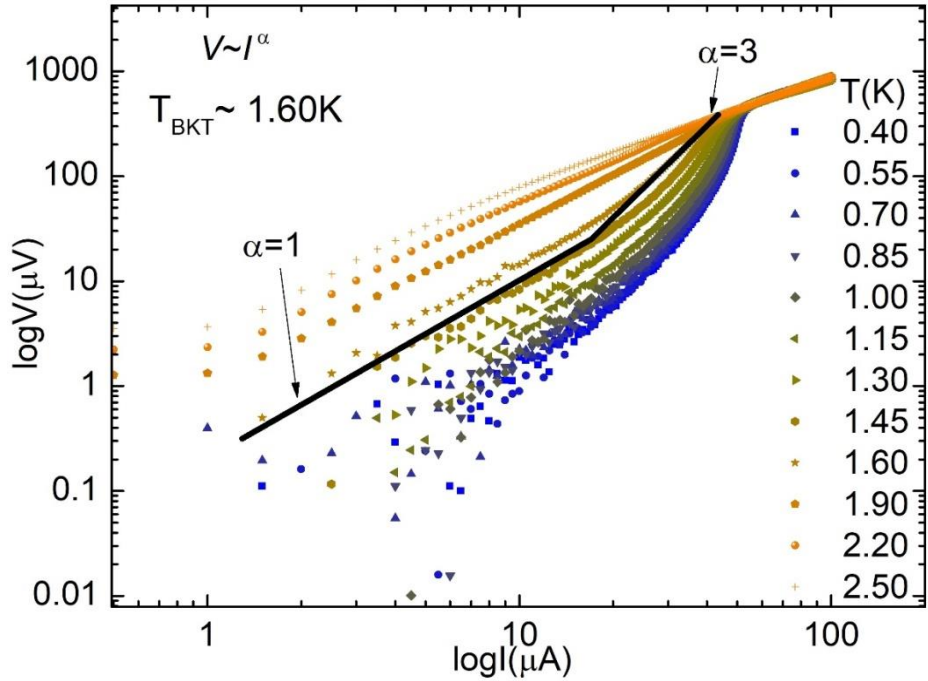

**Figure S4** – *V-I* curves at different temperatures below $T_C$, defined as the onset of the resistive drop (see main text). The two black lines are guides to the eye representing the dependences $V \propto I$ and $V \propto I^3$. Whereas the *V(I)* characteristic behaves in the characteristic nonlinear fashion $V \propto I^\alpha$ near the critical current, the traces ultimately evolve towards linearity at the lowest currents.

## 5. Estimation of the in-plane upper critical field

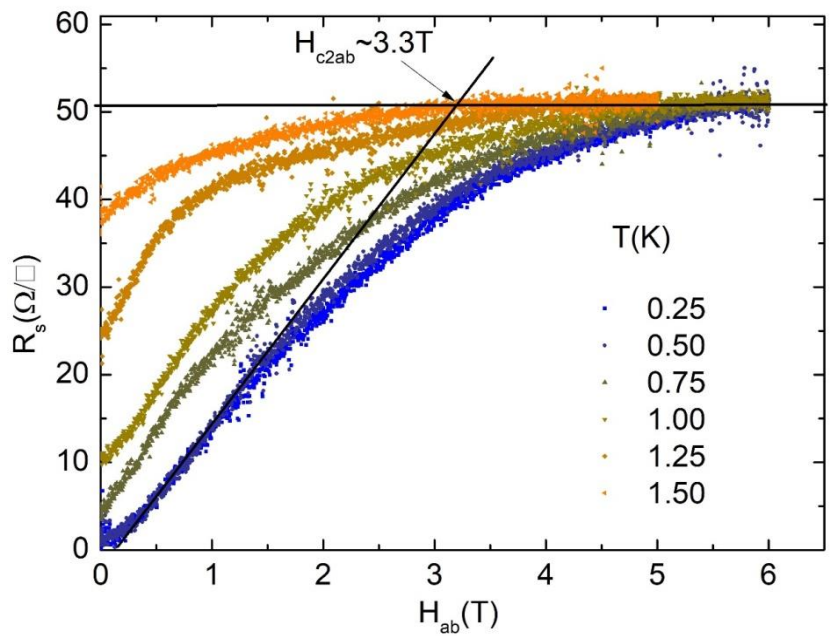

**Figure S5** – Sheet resistance at different values of in-plane magnetic field, for doping level around $4\times10^{14}$ cm$^{-2}$. The in-plane upper critical field $H_{c2,\parallel}$ is estimated by the intersection of the two black lines to be 3.3 T.

The lack of an in-situ rotation stage in our $^3$He refrigerator prevented us from making transport measurements under in-plane field. Therefore, we performed most magnetic-field-dependent measurements with the field normal to the sample plane. To investigate the response under in-plane field, the vacuum chamber had be open, the sample probe removed, rotated to the appropriate orientation, and reloaded into the cryostat. After this procedure, the sample quality was seen to degrade to some extent, which translates into the noise seen in the traces of Fig. S5.

## 6. Reproducibility across different devices and densities

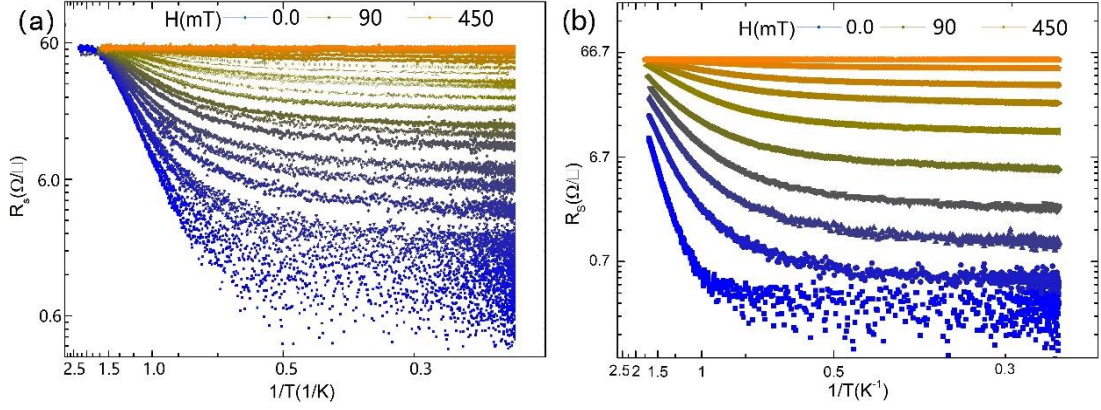

**Figure S6** – Sheet resistance at different values of out-of-plane magnetic field for doping near $3.2\times10^{14}$ cm$^{-2}$ (a) and $4\times10^{14}$ cm$^{-2}$ (b) of another TiSe$_2$ device with similar characteristics to that discussed in the main text. The resistance traces display the same behavior seen in Fig. 3 of the main text, where the leveling of resistance as $T \to 0$ and the positive magnetoresistance at $T \approx 0$ are characteristic of the AQM state. This behavior is seen consistently at different densities within the superconducting dome of each individual device, and also across distinct devices.